\documentclass[aps,amsmath,preprint,epsf,superscriptaddress,nofootinbib]{revtex4-1}
\pdfoutput=1
\usepackage{graphicx}
\usepackage{bm}
\usepackage{color}
\usepackage{amsmath,amssymb}	
\usepackage{hyperref}
\usepackage{setspace}
\begin{document}

\def\a{\alpha}
\def\b{\beta}
\def\c{\varepsilon}
\def\d{\delta}
\def\e{\epsilon}
\def\f{\phi}
\def\g{\gamma}
\def\h{\theta}
\def\k{\kappa}
\def\l{\lambda}
\def\m{\mu}
\def\n{\nu}
\def\p{\psi}
\def\q{\partial}
\def\r{\rho}
\def\s{\sigma}
\def\t{\tau}
\def\u{\upsilon}
\def\v{\varphi}
\def\w{\omega}
\def\x{\xi}
\def\y{\eta}
\def\z{\zeta}
\def\D{\Delta}
\def\G{\Gamma}
\def\H{\Theta}
\def\L{\Lambda}
\def\F{\Phi}
\def\P{\Psi}
\def\S{\Sigma}

\def\o{\over}
\def\beq{\begin{eqnarray}}
\def\eeq{\end{eqnarray}}
\newcommand{\gsim}{ \mathop{}_{\textstyle \sim}^{\textstyle >} }
\newcommand{\lsim}{ \mathop{}_{\textstyle \sim}^{\textstyle <} }
\newcommand{\vev}[1]{ \left\langle {#1} \right\rangle }
\newcommand{\bra}[1]{ \langle {#1} | }
\newcommand{\ket}[1]{ | {#1} \rangle }
\newcommand{\EV}{ {\rm eV} }
\newcommand{\KEV}{ {\rm keV} }
\newcommand{\MEV}{ {\rm MeV} }
\newcommand{\GEV}{ {\rm GeV} }
\newcommand{\TEV}{ {\rm TeV} }
\def\diag{\mathop{\rm diag}\nolimits}
\def\Spin{\mathop{\rm Spin}}
\def\SO{\mathop{\rm SO}}
\def\O{\mathop{\rm O}}
\def\SU{\mathop{\rm SU}}
\def\U{\mathop{\rm U}}
\def\Sp{\mathop{\rm Sp}}
\def\SL{\mathop{\rm SL}}
\def\tr{\mathop{\rm tr}}

\def\IJMP{Int.~J.~Mod.~Phys. }
\def\MPL{Mod.~Phys.~Lett. }
\def\NP{Nucl.~Phys. }
\def\PL{Phys.~Lett. }
\def\PR{Phys.~Rev. }
\def\PRL{Phys.~Rev.~Lett. }
\def\PTP{Prog.~Theor.~Phys. }
\def\ZP{Z.~Phys. }

\setlength{\textwidth}{16cm}
\setlength{\textheight}{22cm}
\setlength{\topmargin}{-1.cm}


\baselineskip 0.7cm

\preprint{IPMU16-0116}
\bigskip

\title{Lower limit on the gravitino mass in low-scale gauge mediation with $m_H\simeq 125$\,GeV}

\author{Masahiro Ibe}
\email[e-mail: ]{ibe@icrr.u-tokyo.ac.jp}
\affiliation{Kavli IPMU (WPI), UTIAS, The University of Tokyo, Kashiwa, Chiba 277-8583, Japan}
\affiliation{ICRR, The University of Tokyo, Kashiwa, Chiba 277-8582, Japan}
\author{Tsutomu T. Yanagida}
\email[e-mail: ]{tsutomu.tyanagida@ipmu.jp}
\affiliation{Kavli IPMU (WPI), UTIAS, The University of Tokyo, Kashiwa, Chiba 277-8583, Japan}

\begin{abstract}
We revisit low-scale gauge mediation models in light of recent observations of 
CMB Lensing and Cosmic Shear which put a severe upper limit on the gravitino mass,
$m_{3/2} \lesssim 4.7$\,eV. 
With such a stringent constraint, many  models of low-scale gauge mediation
are excluded when the squark masses are required to be rather large to explain 
the observed Higgs boson mass.
In this note, we discuss a type of low-scale gauge mediation models 
which satisfy both the observed Higgs boson mass and the upper limit on
the gravitino mass.
We also show that the gravitino mass cannot be smaller than about 1\,eV even in such models, 
which may be tested in future observations of 21 cm line fluctuations.
\end{abstract}

\maketitle

\section{Introduction}
Low-scale gauge mediation models with a light gravitino mass, $m_{3/2} < O(10)$\,eV, is very attractive, 
since the gravitino with a mass in this range does not cause astrophysical nor cosmological problems\,\cite{Pagels:1981ke,Moroi:1993mb,Feng:2010ij}.
In particular, such a light gravitino is consistent with high reheating temperature which is essential for many baryogenesis scenarios
as typified in thermal leptogenesis~\cite{Fukugita:1986hr}~\cite[see][for review]{Giudice:2003jh,Buchmuller:2005eh,Davidson:2008bu}.
A small gravitino mass is also  motivated since it may require a milder fine-tuning of the cosmological constant 
due to a smaller supersymmetry (SUSY) breaking scale.

Recently, a  severe upper limit on the mass of the light stable gravitino, $m_{3/2} \lesssim 4.7$\,eV (95\%\,C.L.),
has been put from CMB Lensing and Cosmic Shear~\cite{Osato:2016ixc}.
With such a stringent constraint, many  models of low-scale gauge mediation
are excluded when the squark masses are required to be rather large to explain 
the observed Higgs boson mass, $m_H \simeq 125$\,GeV~\cite{Aad:2015zhl}.
For example, we immediately find that the above upper limit on the gravitino mass excludes
models in which the messenger fields couple to supersymmetry breaking sector 
perturbatively~\cite{Yanagida:2012ef}.

In this short note, we point out that the low-scale gauge mediation model can explain 
the observed Higgs boson mass even for $m_{3/2}\lesssim 4.7$\,eV
when the messenger fields strongly couple to the SUSY breaking sector.
We also show that the gravitino mass cannot be smaller than about 1\,eV even in such models,
which can be tested by future observations of 21\,cm line fluctuations\,\cite{Oyama:2016lor}.

\section{Models with low-scale gauge mediation}
\subsection{Low-scale gauge mediation and Higgs boson mass}
In this note, we are interested in models with a gravitino mass in the eV range. 
For such a light gravitino mass, the SUSY breaking scale must be low as,
\begin{eqnarray}
\label{eq:Fgravitino}
\sqrt{F} \sim 65\,{\rm TeV} \times \left( \frac{m_{3/2}}{1\,{\rm eV}}\right)^{1/2}\ .
\end{eqnarray}
Now, let us suppose that there are $N_m$ pairs of 
$\Psi$ and $\bar \Psi$ which are in the fundamental and anti-fundamental representations of 
$SU(5)_{\rm GUT} \supset SU(3)_C \times SU(2)_L \times U(1)_Y$, 
respectively.
The messenger fields couple to a SUSY breaking sector via a superpotential,
\begin{eqnarray}
\label{eq:superpotential}
W = y Z \Psi \bar{\Psi}  + M_m \Psi \bar{\Psi}\ .
\end{eqnarray}
Here, the SUSY breaking sector is encapsulated in $Z$ whose vacuum expectation value (VEV) is assumed to be $\vev{Z} = \theta^2 F$.
The coupling between the messenger fields and the SUSY breaking sector is given by the term proportional to $y$.
In this simple setup, the mass splitting between the messenger scalars and the fermions is given by
\begin{eqnarray}
F_m = y F\ .
\end{eqnarray}
It should be noted that the mass splitting is required to be smaller than the messenger mass scale, $M_m$, i.e.,
\begin{eqnarray}
\label{eq:nontachyonic}
F_m <  M_m^2\ ,
\end{eqnarray}
to avoid the tachyonic messenger fields.

Below the messenger scale, the  masses of superparticles are given by,
\begin{eqnarray}
m_{\rm gaugino} &\simeq& N_m \left( \frac{g^2}{16\pi^2} \right)\frac{yF}{M_{m}}\ ,  \\
m_{\rm sfermion}^2 &\simeq& 2C_2 N_m \left( \frac{g^2}{16\pi^2} \right)^2\left(\frac{yF}{M_{m}}\right)^2\ . 
\end{eqnarray}
Here, $C_2$ is the quadratic Casimir invariant of representations of each sfermion,
and  $g$ represents gauge coupling constant of the minimal SUSY Standard Model (MSSM).
To satisfy the cosmological constraint on the gravitino mass,  $m_{3/2}<4.7$\,eV, the SUSY breaking 
scale is required to be $\sqrt{F}\lesssim 140$\,TeV (see Eq.\,(\ref{eq:Fgravitino})).
By combined with the non-tachyonic messenger condition Eq.\,(\ref{eq:nontachyonic}), we find that the soft terms are limited from above;
\begin{eqnarray}
\label{eq:upper}
m_{\rm gaugino} &\lesssim& N_m \left( \frac{g^2}{16\pi^2} \right)(y F)^{1/2} \simeq 0.9\,{\rm TeV}\times N_m\, y^{1/2} g^2 \left( \frac{m_{3/2}}{4.7\,\rm eV}\right)^{1/2}  \ , \\
m_{\rm sfermion}^2 &\lesssim& 2C_2 N_m \left( \frac{g^2}{16\pi^2} \right)^2y F
\simeq (1.3\,{\rm TeV})^2 \times C_2 N_m g^4 y \left( \frac{m_{3/2}}{4.7\,\rm eV}\right) \ ,
\end{eqnarray}
at the messenger scale.

\begin{figure}[tbp]
	\centering
		\begin{minipage}{.46\linewidth}
  \includegraphics[width=\linewidth]{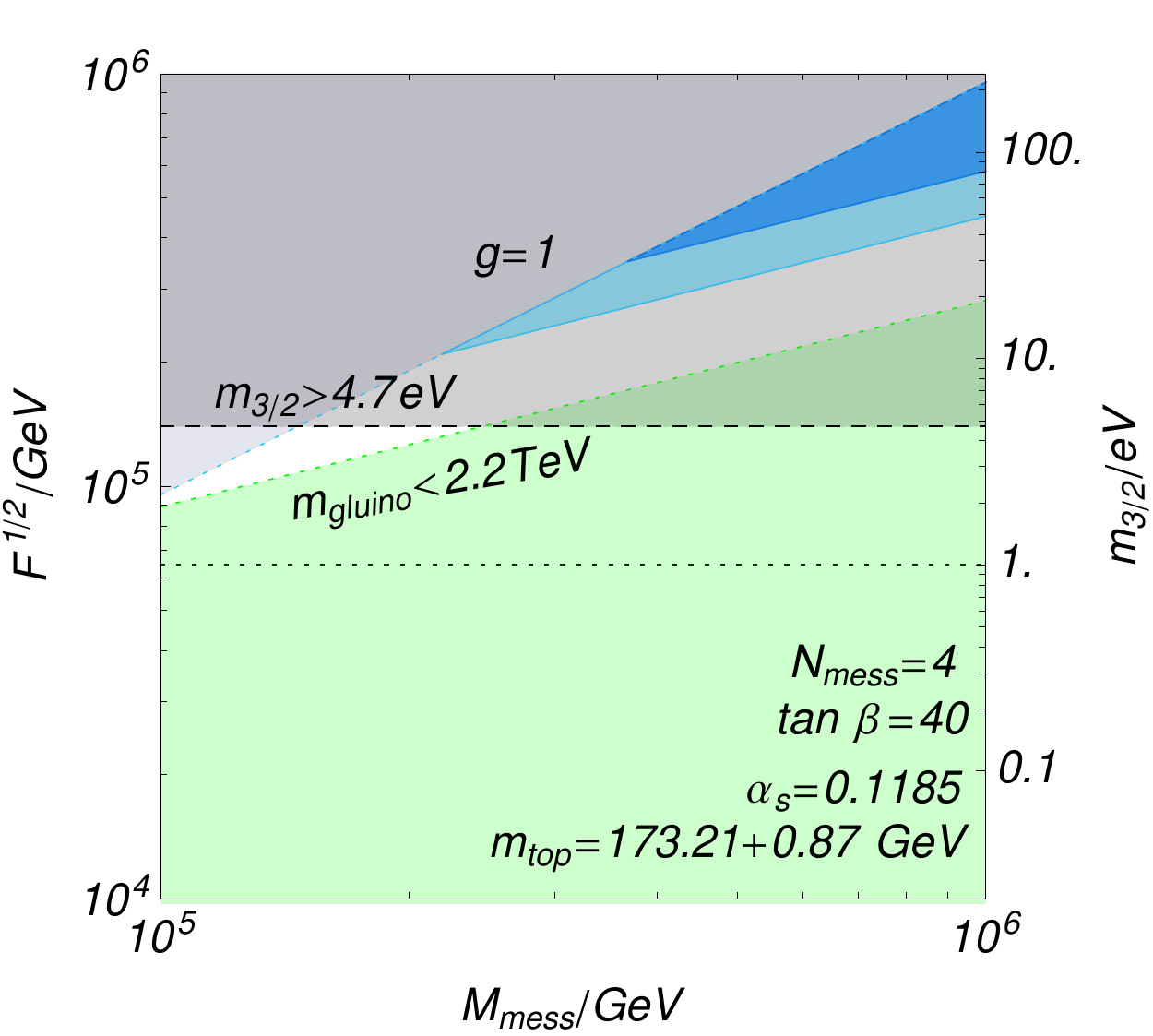}
 \end{minipage}
 \hspace{1cm}
 \begin{minipage}{.46\linewidth}
  \includegraphics[width=\linewidth]{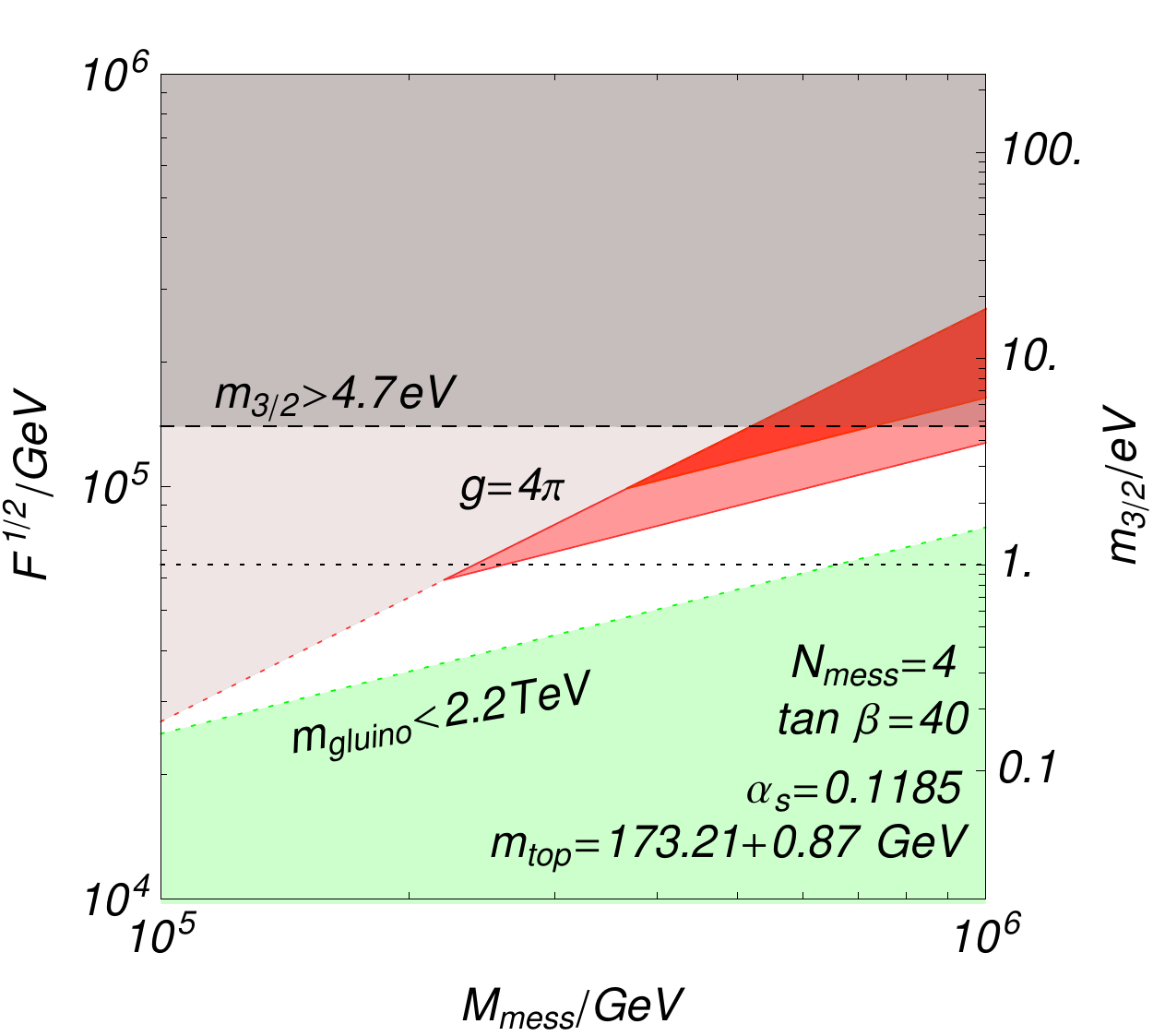}
 \end{minipage}
	\caption{(Left) In the blue shaded region, the observed Higgs boson mass can be explained by the top Yukawa radiative correction
	for $y=1$.
	(Right) In the red shaded region, the observed Higgs boson mass can be explained by the top Yukawa radiative correction
	for $y=4\pi$. 
In both panels, the upper left regions are excluded due to the tachyonic messenger fields, i.e.  $M_m^2 < F_m$.
The regions with $m_{3/2}\gtrsim 4.7$\,eV are excluded by the cosmological constraints.
The green shaded regions are excluded by the null results of searches for the tau slepton  at the LHC.
We take $N_m =4$ and $\tan\beta = 40$, although the results do not depend on $\tan\beta$  significantly
as long as $\tan\beta ={\cal O}(10)$.
	}
	\label{fig:GMSB}
\end{figure}

Let us discuss whether the above soft masses can be consistent with the observed Higgs boson mass, $m_H \simeq 125$\,GeV.
In the MSSM, the Higgs boson mass is constrained as $m_H\lesssim  m_Z$ at the tree level,
which is enhanced by the top Yukawa radiative corrections\,\cite{Okada:1990vk,*Ellis:1990nz,*Haber:1990aw,*Okada:1990gg,*Espinosa:1999zm}.
Then, the observed Higgs boson mass, $m_H \simeq 125$\,GeV, requires the squark masses (in particular the stop masses) 
in multi-TeV range, which is in tension with the squark masses in Eq.\,(\ref{eq:upper}) for $m_{3/2}< 4.7$\,eV.

In Fig.\,\ref{fig:GMSB}, we show the parameter region which is consistent with the observed Higgs boson mass,
$m_H = 125.09\pm0.21 \pm 0.11$\,GeV~\cite{Aad:2015zhl}.
In the figure, the  Higgs boson mass is consistently explained  at the 2$\sigma$ level in the blue and red shaded regions 
for $y = 1$ and $y = 4\pi$, respectively.%
\footnote{In our analysis, we define the $\chi^2$ estimator,
\begin{eqnarray}
\chi^2 = \frac{(m_H-125.09\,{\rm GeV})^2}{(0.21\,{\rm GeV})^2+(0.11\,{\rm GeV})^2 + \delta m_H^2}\ ,
\end{eqnarray}
where $\delta m_H$ denotes the theoretical uncertainty.
}
Here, we take $N_m = 4$, so that the resultant squark masses are as large as possible
for given $M_m$ and $F$ while keeping the perturbativity of the gauge coupling constants 
in the MSSM up to the scale of the grand unification.
In our numerical analysis, we use {\tt softsusy-3.7.3}~\cite{Allanach:2001kg} to 
solve the renormalization group evolution of the MSSM parameters. 
The Higgs boson mass is calculated by {\tt FeynHiggs-2.10.0}~\cite{Heinemeyer:1998yj,*Heinemeyer:1998np,*Degrassi:2002fi,*Frank:2006yh,*Hahn:2013ria}.
To take scheme dependences of the Higgs mass estimations into account, 
we also estimated the Higgs boson mass by using another code {\tt susyHD}~\cite{Vega:2015fna}.
The corresponding parameter regions  are shaded by darker blue/red.
The figures show that the results obtained by using  {\tt susyHD} require  slightly 
higher SUSY breaking scales (and hence heavier squark masses) to achieve the 
observed Higgs boson mass.

As the left panel shows, the region which is consistent with the observed Higgs boson mass
is excluded by the cosmological constraint on the gravitino mass when the 
messenger fields couple to the SUSY breaking sector perturbatively.%
\footnote{Here, we take the upper limit on the top quark mass
$m_{t}= 173.21\pm 0.51\pm 0.71$\,GeV\,\cite{Agashe:2014kda}, which 
leads to a heavier Higgs boson mass for given squark masses.}
When the messenger couple to the SUSY breaking sector strongly, i.e. $y = {\cal O}(4\pi)$,
on the other hand, the observed Higgs boson mass can be explained even for $m_{3/2} \lesssim 4.7$\,eV.
The figure also shows that there is a lower limit on the gravitino mass from the observed Higgs boson mass,
\begin{eqnarray}
m_{3/2} \gtrsim 0.8\,{\rm eV} \times \left(\frac{4\pi}{y}\right)\ .
\end{eqnarray}
According to Ref.\,\cite{Oyama:2016lor}, the gravitino dark matter in this range, i.e. $m_{3/2} \sim 1$\,eV can be 
tested by  future observations of 21\,cm line fluctuations with Square Kilometre Array\,\cite{ska}
and/or Omniscope\,\cite{Tegmark:2008au,2010PhRvD..82j3501T}.

In the figures, the green shaded regions are excluded by SUSY searches at the LHC.
By remembering that the next-to-lightest SUSY particle (NLSP) is the tau slepton for $N_m = 4$,
we show the lower limit on the gluino mass, $m_{\rm gluino} \gtrsim 2.2$--$2.3$\,TeV,
from the null results of searches for the tau slepton~\cite{Aaboud:2016zpr}.%
\footnotemark[3]
The figure shows that the cosmological constraints are more stringent compared with the constraints
put by direct searches at the LHC.

\subsection{Strongly interacting models}
Here, let us illustrate how the strong coupling 
between the messenger sector and the SUSY breaking sector, i.e. $y = {\cal O}(4\pi)$ is achieved.
When these two sectors are strongly interacting, 
the vacuum structure of the SUSY breaking sector is inevitably affected by the coupling to the messenger fields,  and in most cases,  SUSY breaking vacua are destabilized.
To avoid this problem, we need to assume that the messenger fields couple to a secondary SUSY breaking
as realized in models of ``cascade SUSY breaking"~\cite{Ibe:2010jb} 
(see also \cite{Dine:1993yw,Dine:1994vc,Dine:1995ag} for earlier works).
There, a secondary SUSY breaking field $S$ couples to the primary SUSY breaking field $Z$
with $\vev{Z} = F\theta^2$ only through the K\"ahler potential,
\begin{eqnarray}
K = |Z|^2 + |S|^2 + \frac{\kappa}{\Lambda^2} |Z|^2 |S|^2 + \cdots \ ,
\end{eqnarray}
where $\Lambda$ is the dynamical scale of the SUSY breaking sector (see \cite{Ibe:2010jb} for more details).
By using the Naive Dimensional Analysis (NDA)~\cite{Cohen:1997rt,Luty:1997fk}, 
the coefficient $\kappa$ is expected to be of ${\cal O}((4\pi)^2)$ when 
both $Z$ and $S$ take part in the strong dynamics 
with the dynamical scales of ${\cal O}(\Lambda)$.
By using the NDA, the primary SUSY breaking scale is also estimated to be%
\footnotetext[3]{This limit is put by assuming $N_m= 3$ in \cite{Aaboud:2016zpr}.
For $N_m = 4$, the constraint might become slightly more stringent due to a relative 
smallness of the squark mass for $N_m=3$.
When, the NLSP is the lightest neutralino, the lower limit on the gluino mass is slightly 
tighter, $m_{\rm gluino} \gtrsim 1.6$--$1.7$\,TeV, which has been put by the null results of searches for 
the photons with missing energy~\cite{ATLASCollaboration:2016wlb}.
}%
\setcounter{footnote}{3}%
\footnote{If we assume IYIT SUSY breaking model~\cite{Izawa:1996pk,Intriligator:1996pu}, this is achieved when the coupling 
between the gauge singlet and $SP(N_c)$ fundamental quarks are of ${\cal O}(4\pi)$.}
\begin{eqnarray}
F_Z  \sim \frac{\Lambda^2}{4\pi}\ .
\end{eqnarray}
As a result, the term proportional to $\kappa$ leads to a soft SUSY breaking mass of $S$,%
\footnote{Here, we assume $\kappa > 0$ for simplicity, although 
a model in \cite{Ibe:2010jb} is viable even for $\kappa < 0$. }
\begin{eqnarray}
\label{eq:ms}
m_S^2 \simeq -\frac{\kappa}{\Lambda^2} \times F_Z^2 \sim - \Lambda^2 \ .
\end{eqnarray}

Now, let us suppose that $S$ and the messenger fields are composite states of some dynamics so that they
couple in the superpotential
\begin{eqnarray}
\label{eq:super2}
W \simeq \frac{k}{ \Lambda^{n-3}} S^n +  \frac{\lambda}{ \Lambda^{n-3}}S^{n-2} \Psi\bar{\Psi}\  \cdots \ ,
\end{eqnarray}
with $n\ge 3$.%
\footnote{A model with $n=5$ is achieved in \cite{Ibe:2010jb}.} 
By the NDA again, we expect $k = {\cal O}((4\pi)^{n-2})$ and  $\lambda = {\cal O}((4\pi)^{n-2})$. 
Then, the scalar potential of $S$ is roughly given by,
\begin{eqnarray}
V \sim  m_S^2 |S|^2 + \frac{k^2}{ \Lambda^{2n-6}} |S|^{2n-2}\ ,
\end{eqnarray}
which leads to the VEVs of $S$,
\begin{eqnarray}
\vev{S} \sim \frac{ \Lambda}{4\pi}\ ,\quad
\vev{F_S} \sim  \frac{\Lambda^2}{4\pi}\ .
\end{eqnarray}
Putting these VEVs into the superpotential in Eq.\,(\ref{eq:super2}), the messenger fields obtain their masses and the mass splittings,
\begin{eqnarray}
M_m \sim \Lambda, \, \quad F_m \sim \Lambda^2\sim 4\pi F_Z\ ,
\end{eqnarray}
which corresponds to $y = {\cal O}(4\pi)$.
In this way, we can construct a model in which the messenger fields and the SUSY breaking couple 
strongly without causing vacuum stability problem.%
\footnote{See also \cite{Yanagida:2010zz} for another strongly interacting messenger model.}

\subsection{Higgs boson mass beyond the MSSM}
So far, in this note, we have confined ourselves to the MSSM where the Higgs boson mass is explained by the top Yukawa radiative corrections.
Here, let us comment on the extensions of the SUSY standard model which can enhance the Higgs boson mass
without requiring large squark masses.

First, let us consider the so-called NMSSM~\cite[for review]{Maniatis:2009re,*Ellwanger:2009dp} in which a newly introduced singlet field couples to the Higgs doublets in the MSSM.
When the singlet--Higgs coupling is rather large, the observed Higgs boson mass can be explained even for relatively light squarks.
However, in the presence of multiple messenger fields, the upper limit on the singlet--Higgs coupling from the Landau pole problem
is severer than in the models without the messenger fields.
As a result, if we require perturbativity to the NMSSM up to the grand unification scale,
the Higgs boson mass cannot be explained in the models with low-scale gauge mediation (with $y={\cal O}(1)$) 
even for the NMSSM~\cite{Yanagida:2012ef}.%
\footnote{Here, we assume that the NMSSM respects the $Z_2$ symmetry. If we allow $Z_3$ violating terms,
it is possible to explain $m_H\simeq 125$\,GeV without having the Landau pole problem.
In such cases, however, we generically suffer from tadpole problem and fine tuning problems. }

As another example to enhance the Higgs boson mass, it is also possible to introduce vector-like matter fields
coupling to the Higgs 
doublets~\cite{Moroi:1991mg,*Moroi:1992zk,Babu:2008ge,Martin:2009bg,Fukushima:2010pm,Endo:2011mc,Evans:2011uq,Csaki:2012fh,Evans:2012uf,Moroi:2016ztz}.
In those extensions, however, the more the vector-like matters are added,
the severer upper limit on the messenger number from the perturbativity of the MSSM gauge coupling constants is put.
With fewer messenger fields, the sparticle masses are difficult to be above the LHC constraints for $m_{3/2}\lesssim 1$\,eV.

One may also consider the extension of the MSSM with an additional $U(1)$ gauge symmetry.
In fact, the Higgs boson can be enhanced by the associated $D$-term potential of the new $U(1)$ gauge interaction
when the Higgs fields are charged under the symmetry~\cite[for review]{Langacker:2008yv}.
For that purpose, however, we need to require that the soft SUSY breaking masses of $U(1)$ breaking
fields should be of the order of the VEV of $U(1)$.
In view of stringent constraints on $Z'$ gauge bosons put by the LHC searches~\cite{ATLAS:2015nsi,Khachatryan:2015dcf},
the required SUSY breaking mass is at least in a few TeV range.
Since we are assuming gauge mediation, the soft masses of $U(1)$ breaking fields should also be provided by gauge mediation.
As a result, for $F^{1/2}\lesssim 65$\,TeV, it seems not easy to achieve consistent model where $U(1)$ extension
explains the observed Higgs boson mass while keeping the squark masses rather small.

Let us also comment on the models with gauge mediation where the Higgs doublets 
and the messenger fields have small mixings~\cite{Chacko:2001km,*Chacko:2002et,Evans:2011bea,*Evans:2012hg,Abdullah:2012tq,Draper:2011aa,Craig:2012xp}.
In this class of models, a rather large $A$-terms are generated which enhances the Higgs boson mass.
As a result, the observed Higgs boson mass can be explained for the sparticles masses in a few TeV range~\cite{Evans:2011bea,*Evans:2012hg}. 
In such models, however, they generically suffer from instability problem of the 
SUSY breaking vacuum when the messenger scale is close to the SUSY breaking scale~\cite{Hisano:2008sy}.%
\footnote{Details of this type of models in the context of the low-scale gauge mediation will
be discussed elsewhere.}

\section{Conclusions and Discussions}
In this paper, we revisited low-scale gauge mediation models in light of recent observation of 
CMB Lensing and Cosmic Shear which put a severe upper limit on the gravitino mass,
$m_{3/2} \lesssim 4.7$\,eV. 
Such a stringent constraint excludes wide range models of low-scale gauge mediation
when the squark masses are required to be rather large to explain the observed Higgs boson mass.
In this note, we pointed out that strongly interacting low-scale gauge mediation still survives
even if we require that the models satisfy both the observed Higgs boson mass and the upper limit on
the gravitino mass.
We also show that the gravitino mass cannot be smaller than about 1\,eV even when the messenger fields
strongly couple to the SUSY breaking sector.

As an interesting aspect of the strongly coupled low-scale gauge mediation 
it may naturally provide dark matter candidate, the baryonic composite states 
in the SUSY breaking sector or the messenger 
sector~\cite{Dimopoulos:1996gy,Hamaguchi:2007rb,*Hamaguchi:2008rv,*Hamaguchi:2009db,Mardon:2009gw,Fan:2010is,Yanagida:2010zz,Ibe:2010jb}. 
The baryonic composite states are given by higher dimensional operators, and hence, 
they couple to the SM particles very weakly. 
As a result, they are expected to be long lived. 
Furthermore, the annihilation cross section of the baryonic composite states via strong interaction can saturates the unitarity limit,
which requries the dark matter mass of ${\cal O}(100)$\,TeV so that the 
dark matter density can be explained by the thermal freeze-out~\cite{Griest:1989wd}.%
\footnote{For some scenario, the effective annihilation cross section of baryonic composite states can exceed the 
unitarity limit on each partial wave modes~\cite{Harigaya:2016nlg}.}
Therefore, the baryonic composite states of the strong dynamics at around ${\cal O}(100)$\,TeV in the low-scale gauge mediation
naturally explain the observed dark matter density.

Taking the thought one step further, this observation might provide an interesting perspective on the naturalness problem.
Let us consider a distribution of the SUSY breaking scale in the ensemble of vacua (or theories)~\cite{Bousso:2000xa,Susskind:2003kw},
which is expected to be biased  towards a lower scale for a flat universe.
When dark matter is provided as composite states of the strongly coupled low-scale gauge mediation,
the dynamical scale cannot be much smaller than ${\cal O}(100)$\,TeV to avoid the lack of dark matter due to a large 
annihilation cross section.
Thus, the final distribution should have a peak at around ${\cal O}(100)$\,TeV, since the scale lower than ${\cal O}(100)$\,TeV is not habitable.%
\footnote{See e.g.~\cite{Tegmark:2005dy,Hellerman:2005yi} for more on habitable conditions. }
Therefore, in this interpretation, the Higgs boson mass and rather heavy squark masses are outcomes of the cosmological selection on the dark matter 
density~\cite{Yanagida:2010zz}.%
\footnote{See \cite{Harigaya:2015yla} for related discussion on the cosmological selection.}

\begin{acknowledgments}
This work is supported in part by Grants-in-Aid for Scientific Research from the Ministry of Education, Culture, Sports, Science, and Technology (MEXT) KAKENHI, 
Japan, No. 25105011 and No. 15H05889 (M. I.) as well as No. 26104009 (T. T. Y.); Grant-in-Aid No. 26287039 (M. I. and T. T. Y.) and  No. 16H02176 (T. T. Y.) 
 from the Japan Society for the Promotion of Science (JSPS) KAKENHI; and by the World Premier International Research Center Initiative (WPI), MEXT, Japan (M. I., and T. T. Y.).
\end{acknowledgments}
%
%

\end{document}